\documentclass[preprint,superscriptaddress,nofootinbib]{revtex4}
  
  \usepackage{amssymb,amsmath}
  \usepackage{graphicx}
  \begin{document}
  \title{Study of $B_{c}^{\ast}$ ${\to}$ ${\psi}(1S,2S)P$,
    ${\eta}_{c}(1S,2S)P$ weak decays}
  \author{Junfeng Sun}
  \affiliation{Institute of Particle and Nuclear Physics,
              Henan Normal University, Xinxiang 453007, China}
  \author{Yueling Yang}
  \affiliation{Institute of Particle and Nuclear Physics,
              Henan Normal University, Xinxiang 453007, China}
  \author{Na Wang}
  \affiliation{Institute of Particle and Nuclear Physics,
              Henan Normal University, Xinxiang 453007, China}
  \author{Jinshu Huang}
  \affiliation{College of Physics and Electronic Engineering,
              Nanyang Normal University, Nanyang 473061, China}
  \author{Qin Chang}
  \affiliation{Institute of Particle and Nuclear Physics,
              Henan Normal University, Xinxiang 453007, China}
  \begin{abstract}
  Motivated by the potential prospects of the $B_{c}^{\ast}$ meson samples
  at hadron colliders,
  the bottom-changing $B_{c}^{\ast}$ ${\to}$ ${\psi}(1S,2S)P$, ${\eta}_{c}(1S,2S)P$
  weak decays are first studied with the perturbative QCD approach,
  where $P$ $=$ ${\pi}$ and $K$.
  It is found that branching ratio of the CKM-favored $B_{c}^{\ast}$ ${\to}$
  $J/{\psi}{\pi}$ decay is about ${\sim}$ ${\cal O}(10^{-8})$, which might be
  measurable at the future LHC experiments.
  \end{abstract}
   \maketitle

  \section{Introduction}
  \label{sec01}
  The $B_{c}^{\ast}$ meson consists of two heavy quarks with
  different flavor numbers, i.e., $\bar{b}c$ ($b\bar{c}$) for the
  $B_{c}^{{\ast}+}$ ($B_{c}^{{\ast}-}$) meson.
  The $B_{c}^{\ast}$ meson is a spin-triplet ground state
  ($n^{2s+1}L_{J}$ $=$ $1^{3}S_{1}$).
  The $B_{c}^{\ast}$ meson lies below the $BD$ meson pair
  threshold. What is more, the mass difference $m_{B_{c}^{\ast}}$ $-$
  $m_{B_{c}}$ ${\simeq}$ 50 MeV \cite{prd86.094510} is far less
  than the mass of pion.
  So, the $B_{c}^{\ast}$ meson decays via the strong interaction are
  completely forbidden.
  However, the $B_{c}^{\ast}$ meson decays through the
  electromagnetic and weak interactions are allowable within the
  standard model of elementary particles.
  The dominant magnetic dipole (M1) transition,
  $B_{c}^{\ast}$ ${\to}$ $B_{c}{\gamma}$, is strongly suppressed by
  the compact phase spaces, which results in a lifetime
  ${\tau}_{B_{c}^{\ast}}$ ${\sim}$ ${\cal O}(10^{-18}\,{\rm s})$
  \cite{epja52.90}.
  Besides, the $B_{c}^{\ast}$ meson carries explicitly nonzero
  bottom and charm quantum numbers ($B$ $=$ $C$ $=$ ${\pm}1$).
  Hence, the $B_{c}^{\ast}$ meson can decay by means of the
  flavor-changing weak transitions.

  The $B_{c}^{\ast}$ meson weak decays, similar to the $B_{c}$
  meson weak decays \cite{qwg,zpc51,prd49,usp38,epjc60.107,prd77.074013,prd89.114019},
  can be divided into three classes:
  (1) the $b$ quark decay with the $c$ quark as a spectator,
  (2) the $c$ quark decay with the $b$ quark as a spectator,
  and (3) the $b$ and $c$ quarks annihilation into a virtual
  $W^{\pm}$ boson.
  The $B_{c}^{\ast}$ meson has a large mass. In addition,
  both constituent quarks of the $B_{c}^{\ast}$ meson can decay
  individually. Therefore, the $B_{c}^{\ast}$ meson has abundant
  weak decay channels.
  However, the $B_{c}^{\ast}$ meson weak decays have received much
  less attention in the past.
  There is no experimental measurement report \cite{pdg} and few
  theoretical investigations concerned with the $B_{c}^{\ast}$ weak decay.
  Fortunately, with the high luminosity and large production
  cross section of the $B_{c}^{\ast}$ meson \cite{plb364,prd54.4344,epjc38.267,prd72.114009}
  at the running LHC, a huge amount of the $B_{c}^{\ast}$ meson
  data samples would be accumulated. Some of the $B_{c}^{\ast}$
  meson weak decays might be explored in the future.
  The $B_{c}^{\ast}$ meson provides another laboratory to
  study the heavy flavor weak decay.

  In this paper, we will study the nonleptonic $B_{c}^{\ast}$
  ${\to}$ ${\psi}(1S,2S)P$, ${\eta}_{c}(1S,2S)P$ decays with the
  perturbative QCD (pQCD) approach \cite{pqcd1,pqcd2,pqcd3},
  where $P$ $=$ ${\pi}$ and $K$.
  Our motivations are as follows.
  Firstly, due to the development of experimental instruments and technology,
  final states of the charmonium and the charged pion and/or kaon
  are easy to identify experimentally.
  With the advancement of high energy hadron collider experiments,
  the $B_{c}$ ${\to}$ ${\psi}(1S,2S)P$ decays have been observed
  \cite{pdg,prl101.012001,prd87.011101,jhep1501.063,prd90.032009,jhep1605.153,prd92.072007}.
  Due to the production cross section ${\sigma}(B_{c}^{\ast})$ ${\gtrsim}$
  ${\sigma}(B_{c})$ in hadron collisions \cite{plb364,prd54.4344,epjc38.267,prd72.114009},
  hopefully, it is anticipated that the $B_{c}^{\ast}$ ${\to}$ ${\psi}(1S,2S)P$,
  ${\eta}_{c}(1S,2S)P$ decays might be observed experimentally in the future.
  A theoretical study on the $B_{c}^{\ast}$ ${\to}$ ${\psi}(1S,2S)P$,
  ${\eta}_{c}(1S,2S)P$ decays is necessary to provide the future experimental
  investigation with an immediate reference.
  Secondly, due to the relations of ${\sigma}(B_{c}^{\ast})$ ${\gtrsim}$ ${\sigma}(B_{c})$
  and $m_{B_{c}^{\ast}}$ ${\simeq}$ $m_{B_{c}}$ \cite{prd86.094510},
  one possible background for the $B_{c}$ meson decays might come from the $B_{c}^{\ast}$
  meson decays into the same final states.
  Hence, the study of the $B_{c}^{\ast}$ ${\to}$ ${\psi}(1S,2S)P$, ${\eta}_{c}(1S,2S)P$
  decays will provide some useful information for the experimental analysis on
  the $B_{c}$ ${\to}$ ${\psi}(1S,2S)P$, ${\eta}_{c}(1S,2S)P$ decays.
  Thirdly, as it is well known, the Cabibbo-Kobayashi-Maskawa (CKM) matrix element
  ${\vert}V_{cb}{\vert}$ could be determined from the semileptonic
  decays of the $B$ meson to the $D^{(\ast)}$ meson.
  However, there exists a more than $3.0\,{\sigma}$ discrepancy between the values from
  exclusive and inclusive determinations\footnotemark[1] \cite{pdg}.
  \footnotetext[1]{The values of the CKM element ${\vert}V_{cb}{\vert}$
  obtained from inclusive and exclusive determinations are
  ${\vert}V_{cb}{\vert}$ $=$ $(42.2{\pm}0.8){\times}10^{-3}$ and
  $(39.2{\pm}0.7){\times}10^{-3}$, respectively \cite{pdg}.}
  The $B_{c}^{(\ast)}$ ${\to}$ ${\psi}(1S,2S)P$, ${\eta}_{c}(1S,2S)P$ decays are induced
  actually by the $b$ ${\to}$ $c$ (or $\bar{b}$ ${\to}$ $\bar{c}$) transition at the
  quark level. The weak interaction coupling and the decay amplitudes are proportional
  to the CKM matrix element ${\vert}V_{cb}{\vert}$
  [see Eq.(\ref{hamilton}) or Eqs.(\ref{amp-bc-psip})-(\ref{amp-bc-ccp})].
  Hence, the $B_{c}^{(\ast)}$ ${\to}$ ${\psi}(1S,2S)P$, ${\eta}_{c}(1S,2S)P$
  decays, together with nonleptonic $\overline{B}$ ${\to}$ $D^{(\ast)}X$ decays and semileptonic
  $\overline{B}$ ${\to}$ $D^{(\ast)}{\ell}\bar{\nu}$ decays, are expected to give more
  stringent constraints on the CKM matrix element ${\vert}V_{cb}{\vert}$,
  other parameters extracted from the $B$ meson decays,
  and contributions from possible new physics.
  Fourthly, owing to the same dynamical mechanism of the bottom quark
  decay, many phenomenological models used for the $B$ meson decays could,
  in principle, be generalized and applied to the $B_{c}^{\ast}$ meson weak decays.
  The practical applicability and reliability of the pQCD approach
  can be reevaluated with the $B_{c}^{\ast}$ ${\to}$ ${\psi}(1S,2S)P$,
  ${\eta}_{c}(1S,2S)P$ decays.
  Further, the $B_{c}^{\ast}$ ${\to}$ ${\psi}(1S,2S)P$, ${\eta}_{c}(1S,2S)P$
  decays provide an opportunity to study polarization effects
  involved in the vector meson decays.

  This paper is organized as follows.
  The theoretical framework and decay amplitudes with the pQCD approach
  are presented in Section \ref{sec02}.
  Section \ref{sec03} is devoted to the numerical results and discussion.
  The last section is a summary.

  \section{theoretical framework}
  \label{sec02}
  \subsection{The effective Hamiltonian}
  \label{sec0201}
  By means of the operator product expansion and the renormalization
  group (RG) method, the effective Hamiltonian responsible for the
  $B_{c}^{\ast}$ ${\to}$ ${\psi}(1S,2S)P$, ${\eta}_{c}(1S,2S)P$ weak
  decays is expressed in terms of four-quark operators with the
  process-independent couplings of the Wilson coefficients
  $C_{i}$ \cite{9512380},
   \begin{equation}
  {\cal H}_{\rm eff}\, =\, \frac{G_{F}}{\sqrt{2}}
   \sum\limits_{q=d,s} V_{cb}^{\ast} V_{uq} \Big\{
    C_{1}\, ( \bar{b}_{\alpha}\, c_{\alpha} )_{V-A}\,
            ( \bar{u}_{\beta}\, q_{\beta} )_{V-A}
  + C_{2}\, ( \bar{b}_{\alpha}\, c_{\beta} )_{V-A}\,
            ( \bar{u}_{\beta}\, q_{\alpha} )_{V-A} \Big\}
   + {\rm h.c.}
   \label{hamilton},
   \end{equation}
  where the Fermi coupling constant $G_{F}$ ${\simeq}$
  $1.166{\times}10^{-5}\,{\rm GeV}^{-2}$ \cite{pdg};
  current operator $(\bar{q}_{1}\,q_{2})_{V-A}$ $=$
  $\bar{q}_{1}\,{\gamma}_{\mu}(1-{\gamma}_{5})\,q_{2}$;
  ${\alpha}$ and ${\beta}$ are color indices.
  The CKM factor $V_{cb}^{\ast}V_{uq}$
  can be written in terms of the Wolfenstein parameters \cite{pdg},
  i.e.,
   \begin{eqnarray}
   V_{cb}^{\ast} V_{ud} &=&
   A\,{\lambda}^{2}-\frac{1}{2}\,A\,{\lambda}^{4}
   -\frac{1}{8}\,A\,{\lambda}^{6}+{\cal O}({\lambda}^{8}),
   \qquad \text{for } P\, =\, {\pi}
   \label{ckm-pi}; \\
   V_{cb}^{\ast} V_{us} &=&
   A\,{\lambda}^{3}+{\cal O}({\lambda}^{8}),
   \qquad \qquad \qquad \qquad \qquad \text{for } P\, =\, K
   \label{ckm-k}.
   \end{eqnarray}

  The auxiliary parameter ${\mu}$ separates the physical contributions
  into two parts. The hard contributions above the scale ${\mu}$ are
  summarized into the Wilson coefficients $C_{i}(\mu)$.
  Due to the properties of asymptotic freedom of QCD forces, the
  Wilson coefficients are, in principle, computable order by order
  with the RG equation improved perturbation theory as long as the
  scale ${\mu}$ is not too small \cite{9512380}.
  The physical contributions below the scale ${\mu}$ are included in
  the hadronic matrix elements (HME) where the local four-quark operators
  are sandwiched between initial and final states.
  The participating hadrons are bounds states of partons.
  With the participation of the strong interaction in the transition
  from quarks to hadrons, especially, the
  presence of long-distance QCD effects and the entanglement of
  nonperturbative and perturbative contributions, how to properly
  evaluate HME is one of major tasks for a serious phenomenology
  of weak decays of heavy flavor hadrons.

  \subsection{Hadronic matrix elements}
  \label{sec0202}
  Phenomenologically, one has to turn to some approximation or
  assumption for the HME calculation.
  The Lepage-Brodsky approach \cite{prd22} is usually applied
  to a hard scattering process, where a hadron transition matrix
  element is generally written as a convolution integral of
  hadron wave functions reflecting the nonperturbative
  contributions and hard scattering amplitudes containing
  perturbative contributions.
  In order to wipe out the endpoint singularities appearing in
  the collinear approximation \cite{qcdf1,qcdf2,qcdf3} and
  suppress the soft contributions, as it is argued with the pQCD
  approach \cite{pqcd1,pqcd2,pqcd3}, the transverse momentum
  $k_{T}$ of valence quarks should be retained and Sudakov
  factor $e^{-S}$ should be introduced for each of the wave functions.
  Finally, the $B_{c}^{\ast}$ weak decay amplitude can be expressed
  as a multidimensional integral of many parts \cite{pqcd2,pqcd3},
  including the hard effects enclosed by the Wilson coefficients
  $C_{i}$, the heavy quark decay amplitudes ${\cal H}$, and
  the universal wave functions ${\Phi}$,
  \begin{equation}
 {\cal A}\, {\sim}\,
  \sum_{i} {\int} {\prod_j}dk_{j}\,
  C_{i}(t)\,{\cal H}_{i}(t,k_{j})\,{\Phi}_{j}(k_{j})\,e^{-S_{j}}
  \label{hadronic},
  \end{equation}
  where $t$ is a typical scale, $k_{j}$ is the momentum of a valence
  quark.

  \subsection{Kinematic variables}
  \label{sec0203}
\begin{figure}[h]
  \includegraphics[width=0.98\textwidth,bb=85 628 525 715]{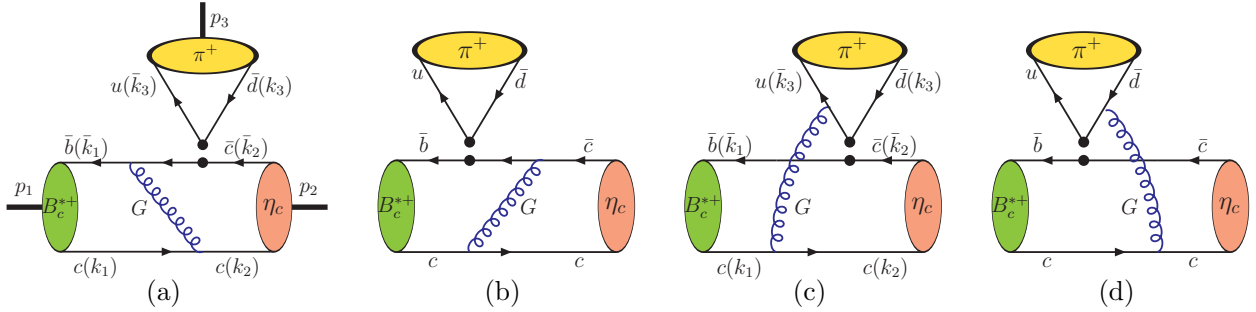}
  \caption{Feynman diagrams for the $B_{c}^{{\ast}+}$ ${\to}$
  ${\eta}_{c}{\pi}^{+}$ decay, where (a,b) are factorizable
  emission topologies, (c,d) are nonfactorizable emission
  topologies.}
  \label{feynman}
  \end{figure}

  In the rest frame of the $B_{c}^{\ast}$ meson, the light-cone
  kinematic variables are defined as follows.
  \begin{equation}
  p_{1}\, =\, \frac{m_{1}}{\sqrt{2}}(1,1,0)
  \label{kine-p1},
  \end{equation}
  \begin{equation}
  p_{2}\, =\, (p_{2}^{+},p_{2}^{-},0)
  \label{kine-p2},
  \end{equation}
  \begin{equation}
  p_{3}\, =\, (p_{3}^{-},p_{3}^{+},0)
  \label{kine-p3},
  \end{equation}
  \begin{equation}
  k_{i}\, =\, x_{i}\,p_{i}+(0,0,\vec{k}_{i{T}})
  \label{kine-ki},
  \end{equation}
  \begin{equation}
  p_{i}^{\pm}\, =\, (E_{i}\,{\pm}\,p)/\sqrt{2}
  \label{kine-pipm},
  \end{equation}
  \begin{equation}
 {\epsilon}_{1}^{{\parallel}}\, =\,\frac{1}{\sqrt{2}}(-1,1,0)
  \label{kine-e1},
  \end{equation}
  \begin{equation}
 {\epsilon}_{2}^{{\parallel}}\, =\, \frac{1}{m_{2}}(p_{2}^{+},-p_{2}^{-},0)
  \label{kine-e2},
  \end{equation}
  \begin{equation}
  {\epsilon}_{1,2}^{\perp}\, =\, (0,0,\vec{1})
  \label{kine-eti},
  \end{equation}
  \begin{equation}
  s\, =\, 2\,p_{2}{\cdot}p_{3}
  \label{kine-s},
  \end{equation}
  \begin{equation}
  t\, =\, 2\,p_{1}{\cdot}p_{2}\, =\ 2\,m_{1}\,E_{2}
  \label{kine-t},
  \end{equation}
  \begin{equation}
  u\, =\, 2\,p_{1}{\cdot}p_{3}\, =\ 2\,m_{1}\,E_{3}
  \label{kine-u},
  \end{equation}
  \begin{equation}
  s\,t +s\,u-t\,u \,=\, 4\,m_{1}^{2}\,p^{2}
  \label{kine-pcm},
  \end{equation}
  where for variables including the momentum $p_{i}$,
  mass $m_{i}$, energy $E_{i}$, and longitudinal (transverse) polarization
  vector ${\epsilon}_{i}^{\parallel}$ (${\epsilon}_{i}^{\perp}$),
  the subscript $i$ $=$ $1$, $2$, $3$ stands for the $B_{c}^{\ast}$
  meson, charmonium ${\psi}$ (${\eta}_{c}$), and pseudoscalar meson $P$,
  respectively;
  $k_{i}$ is the momentum of a valence quark;
  $x_{i}$ and $k_{iT}$ are the longitudinal momentum fraction and
  transverse momentum, respectively;
  $p$ is the center-of-mass momentum of final states;
  $s$, $t$ and $u$ are the Lorentz invariant parameters.
  The kinematic variables are displayed in Fig.\ref{feynman}(a).

  \subsection{Wave functions}
  \label{sec0204}
  It is seen from Eq.(\ref{hadronic}) that wave functions are
  essential input parameters with the pQCD approach.
  And in general, wave functions are universal, i.e., process independent.
  Following the notations in Refs.\cite{prd65,jhep.0605.004},
  wave functions of participating mesons are defined as
  \begin{equation}
 {\langle}0{\vert}\bar{b}_{i}(0)c_{j}(z)
 {\vert}B_{c}^{\ast}(p,{\epsilon}^{\parallel}){\rangle}\,
 =\, \frac{f_{B_{c}^{\ast}}}{4} {\int}d^{4}k\,e^{-ik{\cdot}z}
  \Big\{ \!\!\not{\!\epsilon}^{\parallel}\, \Big[
  m_{B_{c}^{\ast}}\, {\Phi}_{B_{c}^{\ast}}^{v}(k) -
  \!\!\not{p}\, {\Phi}_{B_{c}^{\ast}}^{t}(k) \Big] \Big\}_{ji}
  \label{wf-bc01},
  \end{equation}
  \begin{equation}
 {\langle}0{\vert}\bar{b}_{i}(0)c_{j}(z)
 {\vert}B_{c}^{\ast}(p,{\epsilon}^{\perp}){\rangle}\,
 =\, \frac{f_{B_{c}^{\ast}}}{4} {\int}d^{4}k\,e^{-ik{\cdot}z}
  \Big\{ \!\!\not{\!\epsilon}^{\perp}\, \Big[
  m_{B_{c}^{\ast}}\, {\Phi}_{B_{c}^{\ast}}^{V}(k) -
  \!\!\not{p}\, {\Phi}_{B_{c}^{\ast}}^{T}(k) \Big] \Big\}_{ji}
  \label{wf-bc02},
  \end{equation}
  \begin{equation}
 {\langle}{\psi}(p,{\epsilon}^{\parallel})
 {\vert}c_{i}(0)\bar{c}_{j}(z){\vert}0{\rangle}\,
 =\, \frac{f_{\psi}}{4} {\int}d^{4}k\,e^{+ik{\cdot}z}
  \Big\{ \!\!\not{\!\epsilon}^{\parallel}\, \Big[
  m_{\psi}\, {\Phi}_{\psi}^{v}(k) +
  \!\!\not{p}\, {\Phi}_{\psi}^{t}(k) \Big] \Big\}_{ji}
  \label{wf-ds02},
  \end{equation}
  \begin{equation}
 {\langle}{\psi}(p,{\epsilon}^{\perp}){\vert}c_{i}(0)\bar{c}_{j}(z){\vert}0{\rangle}\,
 =\, \frac{f_{\psi}}{4} {\int}d^{4}k\,e^{+ik{\cdot}z}
  \Big\{ \!\!\not{\!\epsilon}^{\perp}\, \Big[
  m_{\psi}\, {\Phi}_{\psi}^{V}(k) +
  \!\!\not{p}\, {\Phi}_{\psi}^{T}(k) \Big] \Big\}_{ji}
  \label{wf-ds03},
  \end{equation}
  \begin{equation}
 {\langle}{\eta}_{c}(p){\vert}c_{i}(0)\bar{c}_{j}(z){\vert}0{\rangle}\,
 =\, \frac{i\,f_{{\eta}_{c}}}{4}{\int}d^{4}k\,e^{+ik{\cdot}z}\,
  \Big\{ {\gamma}_{5}\, \Big[ \!\!\not{p}\,{\Phi}_{{\eta}_{c}}^{a}(k)
  +m_{{\eta}_{c}}\,{\Phi}_{{\eta}_{c}}^{p}(k) \Big] \Big\}_{ji}
  \label{wf-ds01},
  \end{equation}
  \begin{equation}
 {\langle}P(p){\vert}\bar{u}_{i}(0)q_{j}(z){\vert}0{\rangle}\,
 =\, \frac{i\,f_{P}}{4}{\int}d^{4}k\,e^{+ik{\cdot}z}\,
  \Big\{ {\gamma}_{5}\Big[ \!\!\not{p}\,{\Phi}_{P}^{a}(k) +
 {\mu}_{P}\,{\Phi}_{P}^{p}(k) +
 {\mu}_{P}\,(\not{n}_{+}\!\!\not{n}_{-}-1)\,{\Phi}_{P}^{t}(k)
  \Big] \Big\}_{ji}
  \label{wf-pi},
  \end{equation}
  where $f_{B_{c}^{\ast}}$, $f_{\psi}$, $f_{{\eta}_{c}}$ and $f_{P}$
  are decay constants;
  wave functions of ${\Phi}^{v,T}$ and ${\Phi}^{a}$ are twist-2;
  ${\Phi}^{V,t}$ and ${\Phi}^{p,t}$ are twist-3;
  ${\mu}_{P}$ $=$ $m_{3}^{2}/(m_{u}+m_{q})$ is the chiral parameter;
  $n_{+}$ and $n_{-}$ are the positive and negative null vectors,
  respectively.

  For the light pseudoscalar meson $P$, only the leading twist
  (twist-2) distribution amplitude (DA) ${\phi}_{P}^{a}(x)$ is involved in
  our calculation (see the Appendix). And the normalized DA ${\phi}_{P}^{a}(x)$
  has the following general structure \cite{jhep.0605.004}:
  \begin{equation}
 {\phi}_{P}^{a}(x)\ =\ 6\,x\,\bar{x}\,
  \Big\{ 1+ \sum\limits_{n=1}
   a_{n}^{P}\, C_{n}^{3/2}(t) \Big\}
  \label{da-pi},
  \end{equation}
  where $\bar{x}$ $=$ $1$ $-$ $x$ and $t$ $=$ $x$ $-$ $\bar{x}$;
  the Gegenbauer moment $a_{n}^{P}$ is a nonperturbative parameter.
  The Gegenbauer polynomials are expressed as:
  \begin{equation}
  C_{1}^{3/2}(t) = 3\,t,
  \quad
  C_{2}^{3/2}(t) = \frac{3}{2}\,(5\,t^{2}-1),
  \quad
  {\cdots}
  \label{polynomials}
  \end{equation}

  The $B_{c}^{\ast}$ meson and charmonium ${\psi}$ and ${\eta}_{c}$
  consist of two heavy flavors.
  The motion of the valence quarks in these mesons should be nearly
  nonrelativistic.
  Taking a similar treatment of the nonrelativistic heavy quarkonium system
  \cite{plb751.171,plb752.322,ijmpa31.1650146,npb911.890},
  DAs for the $B_{c}^{\ast}$ meson and charmonium can be written as
   \begin{equation}
  {\phi}_{B_{c}^{\ast}}^{v,T}(x) = A\, x\,\bar{x}\,
  {\exp}\Big\{ -\frac{\bar{x}\,m_{c}^{2}+x\,m_{b}^{2}}
                     {8\,{\omega}_{1}^{2}\,x\,\bar{x}} \Big\}
   \label{da-bc-ev},
   \end{equation}
   \begin{equation}
  {\phi}_{B_{c}^{\ast}}^{t}(x) = B\, (\bar{x}-x)^{2}\,
  {\exp}\Big\{ -\frac{\bar{x}\,m_{c}^{2}+x\,m_{b}^{2}}
                     {8\,{\omega}_{1}^{2}\,x\,\bar{x}} \Big\}
   \label{da-bc-et},
   \end{equation}
   \begin{equation}
  {\phi}_{B_{c}^{\ast}}^{V}(x) = C\, \big\{ 1+(\bar{x}-x)^{2} \big\}\,
  {\exp}\Big\{ -\frac{\bar{x}\,m_{c}^{2}+x\,m_{b}^{2}}
                     {8\,{\omega}_{1}^{2}\,x\,\bar{x}} \Big\}
   \label{da-bc-tv},
   \end{equation}
   \begin{equation}
  {\phi}_{{\eta}_{c}(1S)}^{a}(x) =
  {\phi}_{{\psi}(1S)}^{v,T}(x) = D\, x\,\bar{x}\,
  {\exp}\Big\{ -\frac{ m_{c}^{2} }{ 8\,{\omega}_{2}^{2}\,x\,\bar{x} } \Big\}
   \label{da-j-ev},
   \end{equation}
   \begin{equation}
  {\phi}_{{\psi}(1S)}^{t}(x) = E\, (\bar{x}-x)^{2}\,
  {\exp}\Big\{ -\frac{ m_{c}^{2} }{ 8\,{\omega}_{2}^{2}\,x\,\bar{x} } \Big\}
   \label{da-j-et},
   \end{equation}
   \begin{equation}
  {\phi}_{{\psi}(1S)}^{V}(x) = F\, \big\{ 1+(\bar{x}-x)^{2} \big\}\,
  {\exp}\Big\{ -\frac{ m_{c}^{2} }{ 8\,{\omega}_{2}^{2}\,x\,\bar{x} } \Big\}
   \label{da-j-tv},
   \end{equation}
   \begin{equation}
  {\phi}_{{\eta}_{c}(1S)}^{p}(x)  = G\,
  {\exp}\Big\{ -\frac{ m_{c}^{2} }{ 8\,{\omega}_{2}^{2}\,x\,\bar{x} } \Big\}
   \label{da-etac-p},
   \end{equation}
   \begin{equation}
  {\phi}_{{\psi}(2S)}^{v,t,V,T}(x)  = H\, {\phi}_{{\psi}(1S)}^{v,t,V,T}(x)\,
   \Big\{ 1+\frac{m_{c}^{2}}{2\,{\omega}_{2}^{2}\,x\,\bar{x}} \Big\}
   \label{da-2j-v},
   \end{equation}
   \begin{equation}
  {\phi}_{{\eta}_{c}(2S)}^{a,p}(x)  = I\, {\phi}_{{\eta}_{c}(1S)}^{a,p}(x)\,
   \Big\{ 1+\frac{m_{c}^{2}}{2\,{\omega}_{2}^{2}\,x\,\bar{x}} \Big\}
   \label{da-2j-p},
   \end{equation}
   where parameter ${\omega}_{i}$ ${\simeq}$ $m_{i}\,{\alpha}_{s}(m_{i})$
   determines the average transverse momentum of valence quarks
   according to
   the power counting rules of nonrelativistic QCD effective theory
   \cite{prd46.4052,prd51.1125,rmp77.1423};
   parameters $A$, $B$, $C$, $D$, $E$, $F$, $G$, $H$, $I$ in
   Eqs.(\ref{da-bc-ev}-\ref{da-2j-p}) could be explicitly determined
   with the following normalization conditions,
   \begin{equation}
  {\int}_{0}^{1}dx\,{\phi}_{B_{c}^{\ast}}^{v,t,V,T}(x)=1
   \label{wave-nb},
   \end{equation}
   \begin{equation}
  {\int}_{0}^{1}dx\,{\phi}_{{\psi}}^{v,t,V,T}(x) =1
   \label{wave-nc},
   \end{equation}
   \begin{equation}
  {\int}_{0}^{1}dx\,{\phi}_{{\eta}_{c}}^{a,p}(x) =1
   \label{wave-nd}.
   \end{equation}

\begin{figure}[h]
  \includegraphics[width=0.98\textwidth,bb=75 615 540 720]{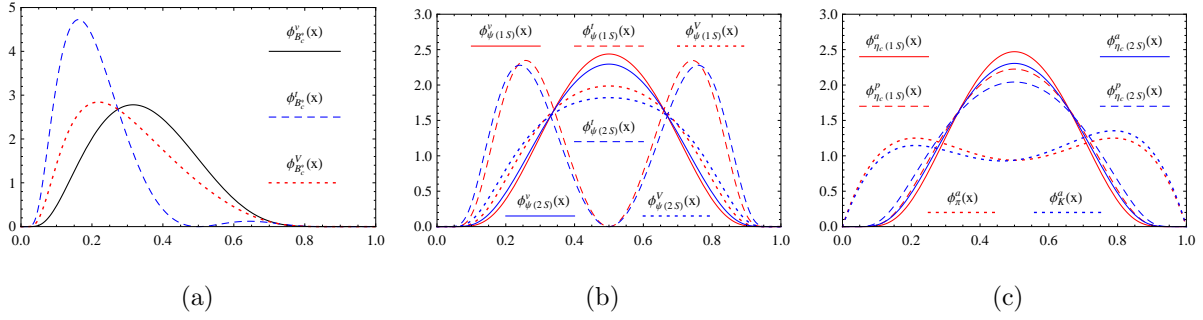}
  \caption{The normalized distribution amplitudes of
  ${\phi}_{B_{c}^{\ast}}^{v,t,V}(x)$, ${\phi}_{\psi}^{v,t,V}(x)$,
  ${\phi}_{{\eta}_{c}}^{a,p}(x)$ and ${\phi}_{{\pi},K}^{a}(x)$.}
  \label{fig:da}
  \end{figure}

  The shape lines of the normalized DAs of participating mesons
  are illustrated in Fig.\ref{fig:da}.
  It is clearly seen from Fig.\ref{fig:da} that (1)
  a broad peak appears at $x$ $<$ $0.5$ region for DAs of the $B_{c}^{\ast}$ meson.
  (2) The shape lines of DAs for the ${\psi}(1S,2S)$ and ${\eta}_{c}(1S,2S)$ mesons
  are symmetric versus $x$, which agree basically with the postulated scenario that
  patrons share momentum fractions according to their masses.
  (3) The differences between DAs of ${\phi}_{\pi}^{a}$ and
  ${\phi}_{K}^{a}$ arise from the flavor symmetry breaking effects
  representing by the Gegenbauer moment $a_{1}^{K}$ ${\neq}$ $0$.
  (4) Owing to the exponential functions, the shape lines of DAs
  in Eqs.(\ref{da-bc-ev}-\ref{da-2j-p}) fall quickly down to zero at
  endpoints $x$, $\bar{x}$ ${\to}$ $0$. So the DAs of
  Eqs.(\ref{da-bc-ev}-\ref{da-2j-p}) will give an effective cut
  for the soft contributions from the endpoints.


  \subsection{Decay amplitudes}
  \label{sec0205}
  The Feynman diagrams for the $B_{c}^{{\ast}}$ ${\to}$ ${\eta}_{c}{\pi}$ decay
  are displayed in Fig.\ref{feynman}, including factorizable topologies
  (a,b) where one gluon couples the $B_{c}^{{\ast}}$ meson with
  the recoiled ${\eta}_{c}$ meson, nonfactorizable topologies (c,d) where
  one gluon is exchanged between the spectator quark and the emitted ${\pi}$
  meson.

  The Lorentz invariant amplitudes for the $B_{c}^{{\ast}}$ ${\to}$ ${\psi}P$,
  ${\eta}_{c}P$ decays are written as
   \begin{equation}
  {\cal A}(B_{c}^{\ast}{\to}{\psi}P)\, =\,
  i\,{\cal F}\,f_{{\psi}}\, \sum\limits_{i}\big\{
  {\cal A}_{i,L}({\epsilon}_{B_{c}^{\ast}}^{{\parallel}},{\epsilon}_{{\psi}}^{{\parallel}})
 +{\cal A}_{i,N}\,({\epsilon}_{B_{c}^{\ast}}^{{\perp}}{\cdot}{\epsilon}_{{\psi}}^{{\perp}})
 +i\,{\cal A}_{i,T}\,{\varepsilon}_{{\alpha}{\beta}{\mu}{\nu}}\,
  p_{B_{c}^{\ast}}^{\alpha}\,p_{{\psi}}^{\beta}\,
  {\epsilon}_{B_{c}^{\ast}}^{{\mu}}\, {\epsilon}_{{\psi}}^{{\nu}} \big\}
   \label{amp-bc-psip},
   \end{equation}
   \begin{equation}
  {\cal A}(B_{c}^{{\ast}}{\to}{\eta}_{c}P)\ =\
  {\cal F}\,f_{{\eta}_{c}}\, \sum\limits_{i}A_{i,P}
   \label{amp-bc-etacp},
   \end{equation}
   \begin{equation}
  {\cal F}\ =\
   \frac{G_{F}}{\sqrt{2}}\, V_{cb}^{\ast}\,V_{uq}\,
   \frac{{\pi}\,C_{F}}{N_{c}}\, f_{B_{c}^{\ast}}\,f_{P}
   \label{amp-bc-ccp},
   \end{equation}
  where the subscript $i$ on ${\cal A}_{i,j}$ corresponds to
  one of the indices in Fig.\ref{feynman}; the subscript $j$ refers
  to different helicity amplitudes; and the expressions of
  building blocks ${\cal A}_{i,j}$ are collected in the Appendix.
  The helicity amplitudes for $B_{c}^{\ast}$ ${\to}$ ${\psi}P$
  decays are defined as
   \begin{equation}
  {\cal M}_{0}\ =\ -{\cal F}\,\sum\limits_{i}
  {\cal A}_{i,L}({\epsilon}_{B_{c}^{\ast}}^{{\parallel}},{\epsilon}_{\psi}^{{\parallel}})
   \label{eq:amp02},
   \end{equation}
   \begin{equation}
  {\cal M}_{\parallel}\ =\ \sqrt{2}\,{\cal F} \sum\limits_{i}
  {\cal A}_{i,N}
   \label{eq:amp03},
   \end{equation}
   \begin{equation}
  {\cal M}_{\perp}\ =\ \sqrt{2}\,{\cal F}\,m_{B_{c}^{\ast}}\,p \sum\limits_{i}
  {\cal A}_{i,T}
   \label{eq:amp04}.
   \end{equation}

  \section{Numerical results and discussion}
  \label{sec03}
  In the rest frame of the $B_{c}^{\ast}$ meson, branching ratios
  for the $B_{c}^{\ast}$ ${\to}$ ${\psi}P$, ${\eta}_{c}P$ decays are
  defined as
   \begin{equation}
  {\cal B}r(B_{c}^{\ast}{\to}{\psi}P)\, =\, \frac{1}{24{\pi}}\,
   \frac{p}{m_{B_{c}^{\ast}}^{2}{\Gamma}_{B_{c}^{\ast}}}\,
   \big\{ {\vert}{\cal M}_{0}{\vert}^{2}
  + {\vert}{\cal M}_{\parallel}{\vert}^{2}
  + {\vert}{\cal M}_{\perp}{\vert}^{2}  \big\}
   \label{br01},
   \end{equation}
   \begin{equation}
  {\cal B}r(B_{c}^{\ast}{\to}{\eta}_{c}P)\, =\, \frac{1}{24{\pi}}\,
   \frac{p}{m_{B_{c}^{\ast}}^{2}{\Gamma}_{B_{c}^{\ast}}}\,
  {\vert}{\cal A}(B_{c}^{\ast}{\to}{\eta}_{c}P){\vert}^{2}
   \label{br02},
   \end{equation}
  where ${\Gamma}_{B_{c}^{\ast}}$ is the full width of
  the $B_{c}^{\ast}$ meson.

   \begin{table}[ht]
   \caption{The numerical values of input parameters.}
   \label{tab:input}
   \begin{ruledtabular}
   \begin{tabular}{lll}
    CKM parameter\cite{pdg}
  & $A$ $=$ $0.811{\pm}0.026$,
  & ${\lambda}$  $=$ $0.22506{\pm}0.00050$, \\ \hline
    $m_{B_{c}^{\ast}}$ $=$ $6332{\pm}9$ MeV\footnotemark[1] \cite{prd86.094510},
  & $m_{{\psi}(1S)}$ $=$ $3096.900{\pm}0.006$ MeV \cite{pdg},
  & ${\Lambda}^{(5)}_{\rm QCD}$ $=$ $210{\pm}14$ MeV \cite{pdg}, \\
    $m_{b}$ $=$ $4.78{\pm}0.06$ GeV \cite{pdg},
  & $m_{{\psi}(2S)}$ $=$ $3686.097{\pm}0.025$ MeV \cite{pdg},
  & ${\Lambda}^{(4)}_{\rm QCD}$ $=$ $292{\pm}16$ MeV \cite{pdg}, \\
    $m_{c}$ $=$ $1.67{\pm}0.07$ GeV \cite{pdg},
  & $m_{{\eta}_{c}(1S)}$ $=$ $2983.4{\pm}0.5$ MeV \cite{pdg},
  & $a_{1}^{\pi}$ $=$ $0$ \cite{jhep.0605.004}, \\
    $f_{B_{c}^{\ast}}$ $=$ $422{\pm}13$ MeV \cite{prd91.114509},
  & $m_{{\eta}_{c}(2S)}$ $=$ $3639.2{\pm}1.2$ MeV \cite{pdg},
  & $a_{2}^{\pi}$ (1\,GeV) $=$ $0.25{\pm}0.15$ \cite{jhep.0605.004}, \\
    $f_{\pi}$ $=$ $130.2{\pm}1.7$ MeV \cite{pdg},
  & $m_{\pi}$ $=$ $139.57$ MeV \cite{pdg},
  & $a_{1}^{K}$ (1\,GeV) $=$ $0.06{\pm}0.03$ \cite{jhep.0605.004}, \\
    $f_{K}$ $=$ $155.6{\pm}0.4$ MeV \cite{pdg},
  & $m_{K}$ $=$ $493.677{\pm}0.016$ MeV \cite{pdg},
  & $a_{2}^{K}$ (1\,GeV) $=$ $0.25{\pm}0.15$ \cite{jhep.0605.004}, \\
   \end{tabular}
   \end{ruledtabular}
   \footnotetext[1]{More predictions of the $B_{c}^{\ast}$ meson mass with
   different models can be found in Table II of  Ref.\cite{prd93.074010}.}
   \end{table}

  The numerical values of some input parameters are listed in
  Table \ref{tab:input}. If it is not specified explicitly,
  their central values will be used in the calculation.

  The decay constant $f_{\psi}$ is related with the branching ratio
  for the leptonic decay ${\psi}$ ${\to}$ $e^{+}e^{-}$ through
  the formula \cite{prd74.034001}
   \begin{equation}
  {\cal B}r({\psi}{\to}e^{+}e^{-})\ =\
   \frac{ 4\,{\pi}\,Q_{c}^{2}\,{\alpha}_{\rm em}^{2}\,f_{\psi}^{2} }
        { 3\,m_{\psi}\,{\Gamma}_{\psi} }
   \label{fpsi},
   \end{equation}
  where $Q_{c}$ is the charge of the charm quark in unit of
  ${\vert}e{\vert}$; ${\alpha}_{\rm em}$ is the fine-structure
  constant of the electromagnetic interaction; ${\Gamma}_{\psi}$
  is full decay width of the ${\psi}$ meson.
  From the available experimental data of both
  ${\cal B}r({\psi}(1S){\to}e^{+}e^{-})$ $=$ $(5.971{\pm}0.032)\%$ and
  ${\cal B}r({\psi}(2S){\to}e^{+}e^{-})$ $=$ $(7.89{\pm}0.17){\times}10^{-3}$
  \cite{pdg}, one can obtain $f_{{\psi}(1S)}$ $=$ $(416.2{\pm}7.4)$
  MeV and $f_{{\psi}(2S)}$ $=$ $(294.6{\pm}7.2)$ MeV, respectively.
  The decay constant $f_{{\eta}_{c}}$ can be extracted from
  the branching ratio for the ${\eta}_{c}$ meson decay into two
  photons using the formula \cite{prd74.034001}
   \begin{equation}
  {\cal B}r({\eta}_{c}{\to}{\gamma}{\gamma})\ =\
   \frac{ 4\,{\pi}\,Q_{c}^{4}\,{\alpha}_{\rm em}^{2}\,f_{{\eta}_{c}}^{2} }
        { m_{{\eta}_{c}}\,{\Gamma}_{{\eta}_{c}} }
   \label{feta}.
   \end{equation}
  With the up-to-date data of
  ${\cal B}r({\eta}_{c}(1S){\to}{\gamma}{\gamma})$ $=$
  $(1.59{\pm}0.13){\times}10^{-4}$ and
  ${\cal B}r({\eta}_{c}(2S){\to}{\gamma}{\gamma})$ $=$
  $(1.9{\pm}1.3){\times}10^{-4}$ \cite{pdg}, one can obtain
  $f_{{\eta}_{c}(1S)}$ $=$ $(337.7{\pm}18.2)$
  MeV and $f_{{\eta}_{c}(2S)}$ $=$ $(243.1{\pm}127.4)$ MeV,
  respectively.

  Besides, the full width of the $B_{c}^{\ast}$ meson,
  ${\Gamma}_{B_{c}^{\ast}}$, is also an essential input
  parameter.
  Because the electromagnetic radiation process $B_{c}^{\ast}$
  ${\to}$ $B_{c}{\gamma}$ dominates the $B_{c}^{\ast}$ meson
  decay, an approximation ${\Gamma}_{B_{c}^{\ast}}$ ${\simeq}$
  ${\Gamma}(B_{c}^{\ast}{\to}B_{c}{\gamma})$ will be used here.
  However, unfortunately, the photon from the $B_{c}^{\ast}$
  ${\to}$ $B_{c}{\gamma}$ process is not hard enough, so,
  it is fairly challenging to identify experimentally.
  The information on ${\Gamma}(B_{c}^{\ast}{\to}B_{c}{\gamma})$
  comes mainly from theoretical estimations.
  Theoretically, the partial decay width of the spin-flip M1 transition
  process has the expression \cite{epja52.90},
   \begin{equation}
  {\Gamma}(B_{c}^{\ast}{\to}B_{c}{\gamma})\ =\
   \frac{4}{3}\,{\alpha}_{\rm em}\, k_{\gamma}^{3}\, {\mu}^{2}_{h}
   \label{m1-width},
   \end{equation}
  where $k_{\gamma}$ is the photon momentum in the rest frame of the
  $B_{c}^{\ast}$ meson;
  ${\mu}_{h}$ is the M1 moment of the $B_{c}^{\ast}$ meson.
  There are plenty of theoretical predictions on
  ${\Gamma}(B_{c}^{\ast}{\to}B_{c}{\gamma})$
  with different approaches, such as
  various potential models \cite{prd49.299,prd49.5845,prd51.3613,
  prd60.074006,prd67.014027,prd70.054017,npa699.649,npa714.183,mpla16.1785}.
  However, because of the incomprehension about ${\mu}_{h}$,
  these estimations suffer from large uncertainties,
  ${\Gamma}(B_{c}^{\ast}{\to}B_{c}{\gamma})$ ${\simeq}$
  $20{\sim}80$ eV \cite{prd49.299,prd49.5845,prd51.3613,prd60.074006,
  prd67.014027,prd70.054017,npa699.649,npa714.183,mpla16.1785}
  (see the numbers in Tables 3 and 6 in Ref.\cite{epja52.90}).
  To give a quantitative estimation, a ballpark guess
  ${\Gamma}_{B_{c}^{\ast}}$ ${\simeq}$ $(50{\pm}30)$ eV will be employed
  here for the moment, where an assumed uncertainty is given to be
  marginally consistent with previous results \cite{prd49.299,
  prd49.5845,prd51.3613,prd60.074006,prd67.014027,prd70.054017,
  npa699.649,npa714.183,mpla16.1785}.

   \begin{table}[ht]
   \caption{Branching ratios for the $B_{c}^{\ast}$ ${\to}$ ${\psi}P$,
   ${\eta}_{c}P$ decays, where the theoretical uncertainties come
   from scale $(1{\pm}0.1)t_{i}$, mass $m_{c}$ and $m_{b}$, the CKM
   parameters, and ${\Gamma}_{B_{c}^{\ast}}$, respectively.}
   \label{tab:br}
   \begin{ruledtabular}
   \begin{tabular}{cc|cc}
   final states & branching ratio & final states & branching ratio \\ \hline
     ${\psi}(1S){\pi}$
   & $( 9.16^{+ 0.98+ 1.13+ 0.68+13.74}_{- 0.45- 0.25- 0.65-~3.43}) {\times}10^{ -8}$
   & ${\psi}(2S){\pi}$
   & $( 3.21^{+ 0.39+ 0.26+ 0.24+ 4.81}_{- 0.17- 0.32- 0.23- 1.20}) {\times}10^{ -8}$ \\
     ${\eta}_{c}(1S){\pi}$
   & $( 2.22^{+ 0.28+ 0.06+ 0.16+ 3.32}_{- 0.12- 0.08- 0.16- 0.83}) {\times}10^{ -8}$
   & ${\eta}_{c}(2S){\pi}$
   & $( 4.59^{+ 0.57+ 0.63+ 0.34+ 6.88}_{- 0.25- 0.86- 0.33- 1.72}) {\times}10^{ -9}$ \\
     ${\psi}(1S)K$
   & $( 7.28^{+ 0.80+ 0.57+ 0.58+10.93}_{- 0.36- 0.46- 0.55-~2.73}) {\times}10^{ -9}$
   & ${\psi}(2S)K$
   & $( 2.37^{+ 0.28+ 0.28+ 0.19+ 3.55}_{- 0.13- 0.18- 0.18- 0.89}) {\times}10^{ -9}$ \\
     ${\eta}_{c}(1S)K$
   & $( 1.67^{+ 0.21+ 0.04+ 0.13+ 2.51}_{- 0.09- 0.07- 0.13- 0.63}) {\times}10^{ -9}$
   & ${\eta}_{c}(2S)K$
   & $( 3.42^{+ 0.43+ 0.45+ 0.27+ 5.12}_{- 0.19- 0.66- 0.26- 1.28}) {\times}10^{-10}$ \\
   \end{tabular}
   \end{ruledtabular}
   \end{table}

  Our numerical results are presented in Table \ref{tab:br},
  where the uncertainties come from the typical scale
  $(1{\pm}0.1)t_{i}$, mass $m_{c}$ and $m_{b}$, the CKM
  parameters, and the decay width ${\Gamma}_{B_{c}^{\ast}}$, respectively.
  The followings are some comments.

  (1)
  Because of the hierarchical relations between the CKM matrix elements
  ${\vert}V_{ud}{\vert}$ $>$ ${\vert}V_{us}{\vert}$,
  branching ratios for the $B_{c}^{\ast}$ ${\to}$ ${\psi}K$, ${\eta}_{c}K$
  decays are generally an order of magnitude less than those for
  the $B_{c}^{\ast}$ ${\to}$ ${\psi}{\pi}$, ${\eta}_{c}{\pi}$ decays
  with the same charmonium in the final states, i.e.,
   \begin{equation}
  {\cal B}r(B_{c}^{{\ast}}{\to}{\psi}{\pi})  >
  {\cal B}r(B_{c}^{{\ast}}{\to}{\psi}K)
   \label{r-01-01},
   \end{equation}
   \begin{equation}
  {\cal B}r(B_{c}^{{\ast}}{\to}{\eta}_{c}{\pi})  >
  {\cal B}r(B_{c}^{{\ast}}{\to}{\eta}_{c}K)
   \label{r-01-02}.
   \end{equation}

  Due to the hierarchical relations between decay constants
  $f_{{\psi}(1S)}$ $>$ $f_{{\psi}(2S)}$ and
  $f_{{\eta}_{c}(1S)}$ $>$ $f_{{\eta}_{c}(2S)}$, along with relatively
  compact phase spaces for final ${\psi}(2S)P$, ${\eta}_{c}(2S)P$ states
  with respect to those for the ${\psi}(1S)P$, ${\eta}_{c}(1S)P$ states,
  there are some hierarchical relations, i.e.,
   \begin{equation}
  {\cal B}r(B_{c}^{{\ast}}{\to}{\psi}(1S)P)  >
  {\cal B}r(B_{c}^{{\ast}}{\to}{\psi}(2S)P)
   \label{r-01-03},
   \end{equation}
   \begin{equation}
  {\cal B}r(B_{c}^{{\ast}}{\to}{\eta}_{c}(1S)P)  >
  {\cal B}r(B_{c}^{{\ast}}{\to}{\eta}_{c}(2S)P)
   \label{r-01-04},
   \end{equation}
  for the same final pseudoscalar meson $P$.

  In addition, due to the conservation of angular momentum, there are
  more wave amplitudes contributing to the $B_{c}^{{\ast}}$ ${\to}$
  ${\psi}P$ decays than the only $p$-wave amplitudes contributing to
  the $B_{c}^{{\ast}}$ ${\to}$ ${\eta}_{c}P$ decays.
  So there are some hierarchical relations, i.e.,
   \begin{equation}
  {\cal B}r(B_{c}^{{\ast}}{\to}{\psi}(1S)P)  >
  {\cal B}r(B_{c}^{{\ast}}{\to}{\eta}_{c}(1S)P)
   \label{r-01-05},
   \end{equation}
   \begin{equation}
  {\cal B}r(B_{c}^{{\ast}}{\to}{\psi}(2S)P)  >
  {\cal B}r(B_{c}^{{\ast}}{\to}{\eta}_{c}(2S)P)
   \label{r-01-06},
   \end{equation}
  for the same final pseudoscalar meson $P$.

  (2)
  The branching ratios of the $B_{c}^{\ast}$ ${\to}$ ${\psi}P$, ${\eta}_{c}P$
  decays are several orders of magnitude less than the branching ratios of
  the $B_{c}$ ${\to}$ ${\psi}P$, ${\eta}_{c}P$ decays \cite{epjc60.107,prd77.074013}.
  So, the possible influence from the $B_{c}^{\ast}$ meson decays could be
  safely neglected when the $B_{c}$ ${\to}$ ${\psi}P$, ${\eta}_{c}P$ decays
  are studied experimentally.
  On the other hand, with the improvement of detection ability and analytical
  techniques, rare $B$ decay modes with branching ratio ${\sim}$ ${\cal O}(10^{-8})$,
  such as the $B^{0}$ ${\to}$ $K^{+}K^{-}$ decay \cite{1610.08288},
  can be accessible at the LHCb experiments now.
  The branching ratios of the $B_{c}^{\ast}$ ${\to}$ ${\psi}{\pi}$
  decays can reach up to ${\cal O}(10^{-8})$.
  In addition, according to the estimation of Ref. \cite{prd72.114009},
  the production cross section of the $B_{c}^{\ast}$ meson is about 30 nb
  at LHC. It is promisingly expected to have more than
  $10^{10}$ $B_{c}^{\ast}$ meson samples, corresponding to hundreds of the
  $B_{c}^{\ast}$ ${\to}$ ${\psi}(1S){\pi}$, ${\psi}(2S){\pi}$ decays, per
  ${\rm ab}^{-1}$ data accumulated at LHC.
  The possible background from the $B_{c}$ ${\to}$ ${\psi}{\pi}$ decays might,
  in principle, be excluded from the invariant mass of final states.
  So, even given the detection efficiency, the $B_{c}^{\ast}$ ${\to}$ $J/{\psi}{\pi}$
  decay is also measurable, although very challenging, at the future LHC experiments.

  \begin{figure}[h]
  \includegraphics[width=0.52\textwidth,bb=105 475 505 720]{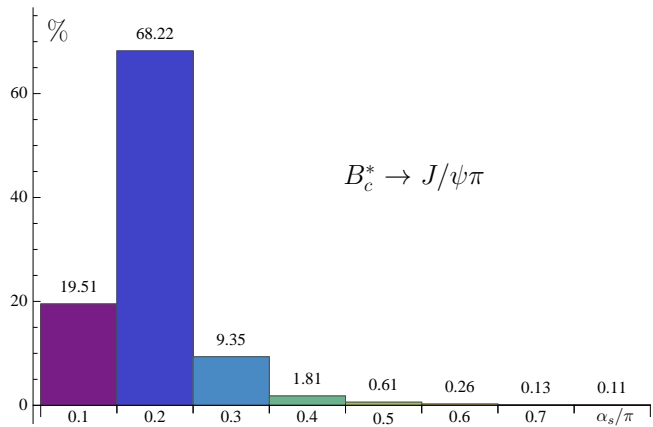}
  \caption{Contributions to branching ratio
  ${\cal B}r(B_{c}^{\ast}{\to}J/{\psi}{\pi})$ from different regions of
  ${\alpha}_{s}/{\pi}$ (abscissa axis), where the numbers above the histogram
  denote the percentage of the corresponding contributions.}
  \label{fig:as}
  \end{figure}

  (3) The spectator quarks in the $B_{c}^{\ast}$ ${\to}$ ${\psi}$ and
  $B_{c}^{\ast}$ ${\to}$ ${\eta}_{c}$ transitions are the heavy charm quark.
  It is usually assumed that
  the charm quark in the $B_{c}^{\ast}$ meson and charmonium might be close to on-shell,
  and the gluons emitted or absorbed by the spectator quarks might be soft.
  It is natural to question the validity of perturbative calculation and
  the practicability of pQCD approach.
  In order to eliminate the doubts, it is necessary to check how many shares come
  from the perturbative domain.
  The contributions to branching ratio ${\cal B}r(B_{c}^{\ast}{\to}J/{\psi}{\pi})$
  from different ${\alpha}_{s}/{\pi}$ region are displayed in Fig.\ref{fig:as}.
  It is clearly seen that more than $85\%$ ($95\%$) contributions to branching
  ratio ${\cal B}r(B_{c}^{\ast}{\to}J/{\psi}{\pi})$ come from the ${\alpha}_{s}/{\pi}$
  ${\le}$ $0.2$ ($0.3$) regions, which implies that the perturbative calculation
  with the pQCD approach is feasible and credible.
  The small Wilson coefficient $C_{1}$ and the small coupling ${\alpha}_{s}$
  at a higher scale ${\mu}$ will account for the small percentage from the
  ${\alpha}_{s}/{\pi}$ ${\le}$ $0.1$ region.
  The tiny share from the ${\alpha}_{s}/{\pi}$ ${\ge}$ $0.5$ region is caused
  by the serious suppression on soft contributions from many factors, such as
  Sudakov factor, DAs for the $B_{c}^{\ast}$ meson and charmonium.
  In addition, a preferable convention to choose the scale as the largest one
  of all virtualities of internal particles [see Eq.(\ref{amp:tab}) and
  Eq.(\ref{amp:tcd})] is employed to ensure the perturbative calculation
  with the pQCD approach.

  (4)
  The theoretical predictions have large uncertainties.
  With the precent predictions on branching ratios of the
  $B_{c}^{\ast}$ ${\to}$ ${\psi}P$, ${\eta}_{c}P$ decays,
  strict constraints on parameters (such as the CKM matrix element
  ${\vert}V_{cb}{\vert}$ and decay width ${\Gamma}_{B_{c}^{\ast}}$)
  cannot be obtained. A global fit with more observables
  seems to be necessary.
  The first uncertainty from the typical scale ${\mu}$ might, in principle,
  be reduced by the inclusion of higher order corrections to HMEs.
  The decay amplitudes are closely related with wave functions
  [see Eq.(\ref{hadronic})], and parameters of $m_{b}$ and $m_{c}$ have much
  influence on wave functions used here.
  The third uncertainty arises mainly from the Wolfenstein parameter $A$,
  i.e., $\frac{2\,{\delta}A}{A}$ ${\sim}$ 6.4\%.
  And a large uncertainty comes from the indefinitive decay width
  ${\Gamma}_{B_{c}^{\ast}}$.
  The uncertainties from $m_{b}$ and $m_{c}$, the CKM parameters,
  and ${\Gamma}_{B_{c}^{\ast}}$ are expected to reduce greatly through
  either the relative ratio of branching ratios or other observables,
  such as the polarization fractions
  $f_{0,{\parallel},{\perp}}$ $=$ $\frac{ {\vert}{\cal M}_{0,{\parallel},{\perp}}{\vert}^{2} }
  { {\vert}{\cal M}_{0}{\vert}^{2}+{\vert}{\cal M}_{\parallel}{\vert}^{2}+{\vert}{\cal M}_{\perp}{\vert}^{2} }$.
  Our studies show that 
  the dominant contributions come from the factorizable topologies.
  There are large cancellations between the nonfactorizable contributions.
  Thus the effects from possible new physics should be imperceptible.
  The more dedicated studies are deserved in the future.

  \section{Summary}
  \label{sec04}
  It is expected that there would be a huge amount of the $B_{c}^{\ast}$ meson
  data samples  at the LHC, and there would be a realistic possibility to search for the
  $B_{c}^{\ast}$ meson weak decays in the future.
  In this paper, the nonleptonic $B_{c}^{\ast}$ ${\to}$
  ${\psi}(1S,2S)P$, ${\eta}_{c}(1S,2S)P$ decays are studied first with a
  phenomenological pQCD approach, in order to offer a ready
  reference for the future experimental analysis.
  It is found that branching ratio for the $B_{c}^{\ast}$ ${\to}$
  $J/{\psi}{\pi}$ decay is about ${\sim}$ ${\cal O}(10^{-8})$,
  which could be accessible at the future experiments.

  \begin{appendix}
  \section{Amplitude building blocks for the $B_{c}^{\ast}$ ${\to}$
           ${\psi}P$, ${\eta}_{c}P$ decays}
  \label{block}
   \begin{eqnarray}
  {\cal A}_{a,P} &=&
  2\,m_{1}\,({\epsilon}_{1}{\cdot}p_{2})\,
  {\int}_{0}^{1}dx_{1}
  {\int}_{0}^{1}dx_{2}
  {\int}_{0}^{\infty}b_{1}db_{1}
  {\int}_{0}^{\infty}b_{2}db_{2}\,
  H_{f}({\alpha},{\beta}_{a},b_{1},b_{2})\,
  {\alpha}_{s}(t_{a})
   \nonumber \\ &{\times}&
  E_{f}(t_{a})\, a_{1}(t_{a})\,
  {\phi}_{B_{c}^{\ast}}^{v}(x_{1})\,
   \Big\{ {\phi}_{{\eta}_{c}}^{a}(x_{2})\,
   ( m_{1}^{2}\,\bar{x}_{2}+m_{3}^{2}\,x_{2} )
   + {\phi}_{{\eta}_{c}}^{p}(x_{2})\, m_{2}\,m_{b} \Big\}
   \label{amp:a-01},
   \end{eqnarray}
   \begin{eqnarray}
  {\cal A}_{a,L} &=&
  {\int}_{0}^{1}dx_{1}
  {\int}_{0}^{1}dx_{2}
  {\int}_{0}^{\infty}b_{1}db_{1}
  {\int}_{0}^{\infty}b_{2}db_{2}\,
   H_{f}({\alpha},{\beta}_{a},b_{1},b_{2})\,
  {\alpha}_{s}(t_{a})\, E_{f}(t_{a})
   \nonumber \\ &{\times}&
  a_{1}(t_{a})\, {\phi}_{B_{c}^{\ast}}^{v}(x_{1})\,
   \Big\{ {\phi}_{\psi}^{v}(x_{2})\,
   ( m_{1}^{2}\,s\,\bar{x}_{2}+m_{3}^{2}\,t\,x_{2} )
  + {\phi}_{\psi}^{t}(x_{2})\, m_{2}\,m_{b}\,u \Big\}
   \label{amp:a-02},
   \end{eqnarray}
   \begin{eqnarray}
  {\cal A}_{a,N} &=& m_{1}\,
  {\int}_{0}^{1}dx_{1}
  {\int}_{0}^{1}dx_{2}
  {\int}_{0}^{\infty}b_{1}db_{1}
  {\int}_{0}^{\infty}b_{2}db_{2}\,
   H_{f}({\alpha},{\beta}_{a},b_{1},b_{2})\,
  {\alpha}_{s}(t_{a})
   \nonumber \\ &{\times}&
  E_{f}(t_{a})\, a_{1}(t_{a})\,
  {\phi}_{B_{c}^{\ast}}^{V}(x_{1})\,
   \Big\{ {\phi}_{\psi}^{V}(x_{2})\, m_{2}\,
   ( u-s\,x_{2} )
  + {\phi}_{\psi}^{T}(x_{2})\, m_{b}\,s \Big\}
   \label{amp:a-03},
   \end{eqnarray}
   \begin{eqnarray}
  {\cal A}_{a,T} &=& 2\,m_{1}\,
  {\int}_{0}^{1}dx_{1}
  {\int}_{0}^{1}dx_{2}
  {\int}_{0}^{\infty}b_{1}db_{1}
  {\int}_{0}^{\infty}b_{2}db_{2}\,
   H_{f}({\alpha},{\beta}_{a},b_{1},b_{2})\,
  {\alpha}_{s}(t_{a})
   \nonumber \\ &{\times}&
  E_{f}(t_{a})\, a_{1}(t_{a})\,
  {\phi}_{B_{c}^{\ast}}^{V}(x_{1})\,
   \Big\{ {\phi}_{\psi}^{V}(x_{2})\, m_{2}\, \bar{x}_{2}
  + {\phi}_{\psi}^{T}(x_{2})\, m_{b} \Big\}
   \label{amp:a-04},
   \end{eqnarray}
   \begin{eqnarray}
  {\cal A}_{b,P} &=&
  2\,m_{1}\,({\epsilon}_{1}{\cdot}p_{2})\,
  {\int}_{0}^{1}dx_{1}
  {\int}_{0}^{1}dx_{2}
  {\int}_{0}^{\infty}b_{1}db_{1}
  {\int}_{0}^{\infty}b_{2}db_{2}\,
  H_{f}({\alpha},{\beta}_{b},b_{2},b_{1})
   \nonumber \\ &{\times}&
 {\alpha}_{s}(t_{b})\, E_{f}(t_{b})\,
  \Big\{ {\phi}_{B_{c}^{\ast}}^{t}(x_{1})\, \Big[
  {\phi}_{{\eta}_{c}}^{p}(x_{2})\, 2\,m_{1}\,m_{2}\,\bar{x}_{1}
 -{\phi}_{{\eta}_{c}}^{a}(x_{2})\,m_{1}\,m_{c} \Big]
   \nonumber \\ &+&
  {\phi}_{B_{c}^{\ast}}^{v}(x_{1})\, \Big[
  {\phi}_{{\eta}_{c}}^{p}(x_{2})\, 2\,m_{2}\,m_{c}
 -{\phi}_{{\eta}_{c}}^{a}(x_{2})\,
  ( m_{2}^{2}\,\bar{x}_{1}+m_{3}^{2}\,x_{1} ) \Big]
   \Big\}\, a_{1}(t_{b})
   \label{amp:b-01},
   \end{eqnarray}
   \begin{eqnarray}
  {\cal A}_{b,L} &=&
  {\int}_{0}^{1}dx_{1}
  {\int}_{0}^{1}dx_{2}
  {\int}_{0}^{\infty}b_{1}db_{1}
  {\int}_{0}^{\infty}b_{2}db_{2}\,
  H_{f}({\alpha},{\beta}_{b},b_{2},b_{1})\,
  {\alpha}_{s}(t_{b})\, E_{f}(t_{b})
   \nonumber \\ &{\times}&
   a_{1}(t_{b})\, {\phi}_{\psi}^{v}(x_{2})\,
   \Big\{ {\phi}_{B_{c}^{\ast}}^{v}(x_{1})\,
  (m_{2}^{2}\,u\,\bar{x}_{1}-m_{3}^{2}\,t\,x_{1})
  + {\phi}_{B_{c}^{\ast}}^{t}(x_{1})\,m_{1}\,m_{c}\,s \Big\}
   \label{amp:b-02},
   \end{eqnarray}
   \begin{eqnarray}
  {\cal A}_{b,N} &=&
  m_{2}\, {\int}_{0}^{1}dx_{1}
  {\int}_{0}^{1}dx_{2}
  {\int}_{0}^{\infty}b_{1}db_{1}
  {\int}_{0}^{\infty}b_{2}db_{2}\,
  H_{f}({\alpha},{\beta}_{b},b_{2},b_{1})\,
  E_{f}(t_{b})
   \nonumber \\ &{\times}&
  {\alpha}_{s}(t_{b})\, a_{1}(t_{b})\, {\phi}_{\psi}^{V}(x_{2})\, \Big\{
    {\phi}_{B_{c}^{\ast}}^{V}(x_{1})\, m_{1}\, (s-u\,x_{1})
  + {\phi}_{B_{c}^{\ast}}^{T}(x_{1})\,m_{c}\,u \Big\}
   \label{amp:b-03},
   \end{eqnarray}
   \begin{eqnarray}
  {\cal A}_{b,T} &=&
  2\, m_{2}\,
  {\int}_{0}^{1}dx_{1}
  {\int}_{0}^{1}dx_{2}
  {\int}_{0}^{\infty}b_{1}db_{1}
  {\int}_{0}^{\infty}b_{2}db_{2}\,
  H_{f}({\alpha},{\beta}_{b},b_{2},b_{1})\,
  {\alpha}_{s}(t_{b})
   \nonumber \\ &{\times}&
   E_{f}(t_{b})\, a_{1}(t_{b})\,
    {\phi}_{\psi}^{V}(x_{2})\, \Big\{
    {\phi}_{B_{c}^{\ast}}^{V}(x_{1})\, m_{1}\, \bar{x}_{1}
  + {\phi}_{B_{c}^{\ast}}^{T}(x_{1})\, m_{c} \Big\}
   \label{amp:b-04},
   \end{eqnarray}
   \begin{eqnarray}
  {\cal A}_{c,P} &=&
   \frac{ 2\,m_{1}\,({\epsilon}_{1}{\cdot}p_{2}) }{ N_{c} }\,
  {\int}_{0}^{1}dx_{1}
  {\int}_{0}^{1}dx_{2}
  {\int}_{0}^{1}dx_{3}
  {\int}_{0}^{\infty}db_{1}
  {\int}_{0}^{\infty}b_{2}db_{2}
  {\int}_{0}^{\infty}b_{3}db_{3}
   \nonumber \\ &{\times}&
   H_{n}({\alpha},{\beta}_{c},b_{1},b_{2},b_{3})\,
   {\alpha}_{s}(t_{c})\,
  \Big\{ {\phi}_{B_{c}^{\ast}}^{v}(x_{1})\,
  {\phi}_{{\eta}_{c}}^{a}(x_{2})\, s\,(x_{2}-\bar{x}_{3})
   \nonumber \\ &+&
  {\phi}_{B_{c}^{\ast}}^{t}(x_{1})\,
  {\phi}_{{\eta}_{c}}^{p}(x_{2})\, m_{1}\, m_{2}\,(x_{1}-x_{2}) \Big\}\,
  {\phi}_{P}^{a}(x_{3})\, E_{n}(t_{c})\, C_{2}(t_{c})
   \label{amp:c-01},
   \end{eqnarray}
   \begin{eqnarray}
  {\cal A}_{c,L} &=&
   \frac{1}{N_{c}}\,
  {\int}_{0}^{1}dx_{1}
  {\int}_{0}^{1}dx_{2}
  {\int}_{0}^{1}dx_{3}
  {\int}_{0}^{\infty}db_{1}
  {\int}_{0}^{\infty}b_{2}db_{2}
  {\int}_{0}^{\infty}b_{3}db_{3}\,
  {\alpha}_{s}(t_{c})\,  E_{n}(t_{c})
   \nonumber \\ &{\times}&
  H_{n}({\alpha},{\beta}_{c},b_{1},b_{2},b_{3})\,
  {\phi}_{P}^{a}(x_{3})\,
   \Big\{ {\phi}_{B_{c}^{\ast}}^{v}(x_{1})\,
  {\phi}_{\psi}^{v}(x_{2})\, 4\,m_{1}^{2}\,p^{2}\,(x_{2}-\bar{x}_{3})
   \nonumber \\ & & \quad +
  {\phi}_{B_{c}^{\ast}}^{t}(x_{1})\, {\phi}_{\psi}^{t}(x_{2})\,
  m_{1}\,m_{2}\,(u\,x_{1}-s\,x_{2}-2\,m_{3}^{2}\,\bar{x}_{3})
   \Big\}\, C_{2}(t_{c})
   \label{amp:c-02},
   \end{eqnarray}
   \begin{eqnarray}
  {\cal A}_{c,N} &=&
   \frac{1}{N_{c}}\,
  {\int}_{0}^{1}dx_{1}
  {\int}_{0}^{1}dx_{2}
  {\int}_{0}^{1}dx_{3}
  {\int}_{0}^{\infty}db_{1}
  {\int}_{0}^{\infty}b_{2}db_{2}
  {\int}_{0}^{\infty}b_{3}db_{3}\,
  H_{n}({\alpha},{\beta}_{c},b_{1},b_{2},b_{3})\,
  {\alpha}_{s}(t_{c})
   \nonumber \\ &{\times}&
  E_{n}(t_{c})\, C_{2}(t_{c})\, {\phi}_{B_{c}^{\ast}}^{T}(x_{1})\,
  {\phi}_{\psi}^{T}(x_{2})\, {\phi}_{P}^{a}(x_{3})\,
  \Big\{ m_{1}^{2}\,s\,(x_{1}-\bar{x}_{3})
  + m_{2}^{2}\,u\,(\bar{x}_{3}-x_{2}) \Big\}
   \label{amp:c-03},
   \end{eqnarray}
   \begin{eqnarray}
  {\cal A}_{c,T} &=&
   \frac{2}{N_{c}}\,
  {\int}_{0}^{1}dx_{1}
  {\int}_{0}^{1}dx_{2}
  {\int}_{0}^{1}dx_{3}
  {\int}_{0}^{\infty}db_{1}
  {\int}_{0}^{\infty}b_{2}db_{2}
  {\int}_{0}^{\infty}b_{3}db_{3}\,
  H_{n}({\alpha},{\beta}_{c},b_{1},b_{2},b_{3})\,
  {\alpha}_{s}(t_{c})
   \nonumber \\ &{\times}&
  E_{n}(t_{c})\, C_{2}(t_{c})\, {\phi}_{B_{c}^{\ast}}^{T}(x_{1})\,
  {\phi}_{\psi}^{T}(x_{2})\, {\phi}_{P}^{a}(x_{3})\,
  \Big\{ m_{1}^{2}\,(x_{1}-\bar{x}_{3})
  + m_{2}^{2}\,(\bar{x}_{3}-x_{2}) \Big\}
   \label{amp:c-04},
   \end{eqnarray}
   \begin{eqnarray}
  {\cal A}_{d,P} &=&
   \frac{ 2\,m_{1}\,({\epsilon}_{1}{\cdot}p_{2}) }{ N_{c} }\,
  {\int}_{0}^{1}dx_{1}
  {\int}_{0}^{1}dx_{2}
  {\int}_{0}^{1}dx_{3}
  {\int}_{0}^{\infty}db_{1}
  {\int}_{0}^{\infty}b_{2}db_{2}
  {\int}_{0}^{\infty}b_{3}db_{3}
   \nonumber \\ &{\times}&
  H_{n}({\alpha},{\beta}_{d},b_{1},b_{2},b_{3})\,
  E_{n}(t_{d})\, \Big\{ {\phi}_{B_{c}^{\ast}}^{t}(x_{1})\,
  {\phi}_{{\eta}_{c}}^{p}(x_{2})\, m_{1}\,m_{2}\,(x_{1}-x_{2})
   \nonumber \\ &+&
  {\phi}_{B_{c}^{\ast}}^{v}(x_{1})\, {\phi}_{{\eta}_{c}}^{a}(x_{2})\,
   (2\,m_{2}^{2}\,x_{2}+s\,x_{3}-t\,x_{1}) \Big\}\,
  {\alpha}_{s}(t_{d})\, C_{2}(t_{d})\, {\phi}_{P}^{a}(x_{3})
   \label{amp:d-01},
   \end{eqnarray}
   \begin{eqnarray}
  {\cal A}_{d,L} &=&
   \frac{1}{N_{c}}\,
  {\int}_{0}^{1}dx_{1}
  {\int}_{0}^{1}dx_{2}
  {\int}_{0}^{1}dx_{3}
  {\int}_{0}^{\infty}db_{1}
  {\int}_{0}^{\infty}b_{2}db_{2}
  {\int}_{0}^{\infty}b_{3}db_{3}\,
  {\alpha}_{s}(t_{d})\, E_{n}(t_{d})
   \nonumber \\ &{\times}&
  H_{n}({\alpha},{\beta}_{d},b_{1},b_{2},b_{3})\,
  {\phi}_{P}^{a}(x_{3})\, \Big\{
  {\phi}_{B_{c}^{\ast}}^{v}(x_{1})\, {\phi}_{\psi}^{v}(x_{2})\,
  4\,m_{1}^{2}\,p^{2}\,(x_{3}-x_{1})
  \nonumber \\ & & \quad -
 {\phi}_{B_{c}^{\ast}}^{t}(x_{1})\, {\phi}_{\psi}^{t}(x_{2})\,
  m_{1}\,m_{2}\,(u\,x_{1}-s\,x_{2}-2\,m_{3}^{2}\,x_{3})
   \Big\}\, C_{2}(t_{d})
   \label{amp:d-02},
   \end{eqnarray}
   \begin{eqnarray}
  {\cal A}_{d,N} &=&
   \frac{1}{N_{c}}\,
  {\int}_{0}^{1}dx_{1}
  {\int}_{0}^{1}dx_{2}
  {\int}_{0}^{1}dx_{3}
  {\int}_{0}^{\infty}db_{1}
  {\int}_{0}^{\infty}b_{2}db_{2}
  {\int}_{0}^{\infty}b_{3}db_{3}\,
  H_{n}({\alpha},{\beta}_{d},b_{1},b_{2},b_{3})\,
  {\alpha}_{s}(t_{d})
   \nonumber \\ &{\times}&
  E_{n}(t_{d})\, C_{2}(t_{d})\, {\phi}_{B_{c}^{\ast}}^{T}(x_{1})\,
  {\phi}_{\psi}^{T}(x_{2})\, {\phi}_{P}^{a}(x_{3})\,
  \Big\{ m_{1}^{2}\,s\,(x_{3}-x_{1})
  + m_{2}^{2}\,u\,(x_{2}-x_{3}) \Big\}
   \label{amp:d-03},
   \end{eqnarray}
   \begin{eqnarray}
  {\cal A}_{d,T} &=&
   \frac{2}{N_{c}}\,
  {\int}_{0}^{1}dx_{1}
  {\int}_{0}^{1}dx_{2}
  {\int}_{0}^{1}dx_{3}
  {\int}_{0}^{\infty}db_{1}
  {\int}_{0}^{\infty}b_{2}db_{2}
  {\int}_{0}^{\infty}b_{3}db_{3}\,
  H_{n}({\alpha},{\beta}_{d},b_{1},b_{2},b_{3})\,
  {\alpha}_{s}(t_{d})
   \nonumber \\ &{\times}&
  E_{n}(t_{d})\, C_{2}(t_{d})\,
  {\phi}_{B_{c}^{\ast}}^{T}(x_{1})\,
  {\phi}_{\psi}^{T}(x_{2})\, {\phi}_{P}^{a}(x_{3})\,
  \Big\{ m_{1}^{2}\,(x_{3}-x_{1})
  + m_{2}^{2}\,(x_{2}-x_{3}) \Big\}
   \label{amp:d-04},
   \end{eqnarray}
  where the subscript $i$ of ${\cal A}_{i,j}$ corresponds to
  the indices of Fig.\ref{feynman}; the subscript $j$ refers
  to possible helicity amplitudes.

  The function $H_{f,n}$ and Sudakov factor $E_{f,n}$ are defined as
   \begin{equation}
   H_{f}({\alpha},{\beta},b_{i},b_{j})\, =\,
   K_{0}(b_{i}\sqrt{-{\alpha}})\, \Big\{
   {\theta}(b_{i}-b_{j}) K_{0}(b_{i}\sqrt{-{\beta}})\,
   I_{0}(b_{j}\sqrt{-{\beta}})
   + (b_{i} {\leftrightarrow} b_{j}) \Big\}
   \label{amp:hf},
   \end{equation}
   \begin{eqnarray}
  H_{n}({\alpha},{\beta},b_{1},b_{2},b_{3}) &=&
  \Big\{ {\theta}(-{\beta})\, K_{0}(b_{3}\sqrt{-{\beta}})
  +\frac{{\pi}}{2}\,
  {\theta}({\beta})\, \Big[ i\,J_{0}(b_{3}\sqrt{{\beta}})
   - Y_{0}(b_{3}\sqrt{{\beta}}) \Big] \Big\}
   \nonumber \\ && \!\!\!\! \!\!\!\! \!\!\!\! \!\!\!\! \!\!\!\!  {\times}\,
   \Big\{ {\theta}(b_{2}-b_{3})\, K_{0}(b_{2}\sqrt{-{\alpha}})\,
   I_{0}(b_{3}\sqrt{-{\alpha}}) + (b_{2} {\leftrightarrow} b_{3})
   \Big\}\, {\delta}(b_{1}-b_{2})
   \label{amp:hn},
   \end{eqnarray}
   \begin{equation}
   E_{f}(t)\ =\ {\exp}\{ -S_{B_{c}^{\ast}}(t)-S_{{\psi},{\eta}_{c}}(t) \}
   \label{sudakov-ef},
   \end{equation}
   \begin{equation}
   E_{n}(t)\ =\ {\exp}\{ -S_{B_{c}^{\ast}}(t)-S_{{\psi},{\eta}_{c}}(t)-S_{P}(t) \}
   \label{sudakov-n},
   \end{equation}
  \begin{equation}
  S_{i}(t)\, =\, s(x_{i},b_{i},p_{i}^{+}) + s(\bar{x}_{i},b_{i},p_{i}^{+})
  +2\,{\int}_{1/b_{i}}^{t}\frac{d{\mu}}{\mu}{\gamma}_{q}
   \label{sudakov-cm},
   \end{equation}
  where $I_{0}$, $J_{0}$, $K_{0}$ and $Y_{0}$ are Bessel functions;
  ${\gamma}_{q}$ $=$ $-{\alpha}_{s}/{\pi}$ is the quark anomalous
  dimension; the expression of $s(x,b,Q)$ can be found in of
  Ref.\cite{pqcd1};
  ${\alpha}$ and ${\beta}_{i}$ are virtualities of gluon and quarks.
  The definitions of the particle virtuality and typical
  scale $t_{i}$ are given as follows.
   \begin{equation}
  {\alpha}\ =\ x_{1}^{2}\,m_{1}^{2}+x_{2}^{2}\,m_{2}^{2}-x_{1}\,x_{2}\,t
   \label{amp:ae},
   \end{equation}
   \begin{equation}
  {\beta}_{a}\ =\ m_{1}^{2} + x_{2}^{2}\,m_{2}^{2}-x_{2}\,t-m_{b}^{2}
   \label{amp:ba},
   \end{equation}
   \begin{equation}
  {\beta}_{b}\ =\ m_{2}^{2}+x_{1}^{2}\,m_{1}^{2}-x_{1}\,t - m_{c}^{2}
   \label{amp:bb},
   \end{equation}
   \begin{equation}
  {\beta}_{c}\ =\ {\alpha} + \bar{x}_{3}^{2}\,m_{3}^{2}
  -x_{1}\,\bar{x}_{3}\,u +x_{2}\,\bar{x}_{3}\,s
   \label{amp:bc},
   \end{equation}
   \begin{equation}
  {\beta}_{d}\ =\ {\alpha}+x_{3}^{2}\,m_{3}^{2}-x_{1}\,x_{3}\,u+x_{2}\,x_{3}\,s
   \label{amp:bd},
   \end{equation}
   \begin{equation}
   t_{a,b}\ =\ {\max}\{ \sqrt{-{\alpha}},\sqrt{{\vert}{\beta}_{a,b}{\vert}},1/b_{1},1/b_{2} \}
   \label{amp:tab},
   \end{equation}
   \begin{equation}
   t_{c,d}\ =\ {\max}\{ \sqrt{-{\alpha}},\sqrt{{\vert}{\beta}_{c,d}{\vert}},1/b_{2},1/b_{3} \}
   \label{amp:tcd}.
   \end{equation}
  \end{appendix}

  
  \end{document}